\documentclass[conference]{IEEEtran}

\IEEEoverridecommandlockouts
\usepackage[sort&compress,numbers]{natbib}
\usepackage{amsmath,amssymb,amsfonts}
\usepackage{graphicx, subcaption}
\usepackage{textcomp}
\usepackage{xcolor}
\usepackage{balance}
\usepackage{comment}
\usepackage{multirow}
\usepackage{siunitx}    
\usepackage{soul}
\usepackage{textcomp}
\usepackage{tabularx}
\usepackage{algorithm}
\usepackage{algpseudocode}
\usepackage{xspace}
\usepackage[draft]{todonotes}
\usepackage{listings}
\usepackage{styles/c_custom_style}
\usepackage{xparse}
\usepackage{courier}

\def\BibTeX{{\rm B\kern-.05em{\sc i\kern-.025em b}\kern-.08em
    T\kern-.1667em\lower.7ex\hbox{E}\kern-.125emX}}
    
\NewDocumentCommand{\codeword}{v}{%
\texttt{\textcolor{black}{#1}}%
}

\usepackage{styles/c_custom_style}
\usepackage{styles/nasm_custom_style}
\newcommand{\shortname}{{\bf MGB}\xspace}
\newcommand{\sa}{{\bf{\it SA}}\xspace}
\newcommand{\cg}{{\bf{\it CG}}\xspace}

\begin{document}

\title{Effective GPU Sharing Under Compiler Guidance}
\author{
\IEEEauthorblockN{Chao Chen\textsuperscript{\textsection}}
\IEEEauthorblockA{\textit{Georgia Institute of Technology} \\
Atlanta, GA, USA \\
chao.chen@gatech.edu}
\and
\IEEEauthorblockN{Chris Porter\textsuperscript{\textsection}}
\IEEEauthorblockA{\textit{Georgia Institute of Technology} \\
Atlanta, GA, USA \\
porter@gatech.edu}
\and
\IEEEauthorblockN{Santosh Pande}
\IEEEauthorblockA{\textit{Georgia Institute of Technology} \\
Atlanta, GA, USA \\
santosh.pande@cc.gatech.edu}
}

\maketitle
\IEEEpeerreviewmaketitle
\begingroup\renewcommand\thefootnote{\textsection}
\footnotetext{These authors contributed equally to the work.}
\endgroup
\thispagestyle{plain}
\pagestyle{plain}

\begin{abstract}
Modern computing platforms tend to deploy multiple GPUs (2, 4, or more) 
on a single node to boost system performance, with each GPU having a large 
capacity in terms of global memory and streaming multiprocessors (SMs).
GPUs are an expensive resource, and boosting utilization of GPUs without 
causing performance degradation of individual workloads is an important 
and challenging problem to be solved. Although services such as MPS provide 
the support for simultaneously executing multiple co-operative kernels on a 
single device, they do not solve the above problem for uncooperative\footnote{Independent kernels arising out of non-MPI processes} kernels, 
MPS being oblivious to the resource needs of each kernel. 

To tackle this problem, 
we propose a \emph{fully automated compiler-assisted scheduling framework}. 
The compiler constructs GPU tasks by identifying kernel launches and 
their related GPU operations (e.g. memory allocations). For each GPU task, a 
probe is instrumented in the host-side code right before its launch point. 
At runtime, the probe conveys the information about the task's resource 
requirements (e.g. memory and compute cores) to a scheduler, such that the 
scheduler can place the task on an appropriate device based on the task's 
resource requirements and devices' load in a memory-safe, resource-aware 
manner. To demonstrate its advantages, we prototyped a throughput-oriented 
scheduler based on the framework, and evaluated it with the Rodinia benchmark 
suite and the Darknet neural network framework on NVIDIA GPUs. The results show 
that the proposed solution outperforms existing state-of-the-art
solutions by leveraging its knowledge about applications'
multiple resource requirements, which include memory as well as SMs.
It improves throughput by up to $2.5\times$ for Rodinia benchmarks, 
and up to $2.7\times$ for Darknet neural networks. In addition, it 
improves job turnaround time by up to $4.9\times$, and limits individual 
kernel performance degradation to at most $2.5\%$.

\end{abstract}

\section{Introduction}
General-purpose graphics processing units (GPGPUs) have become essential 
components in modern data centers and high-performance computing (HPC) 
systems. They provide the massive computing capacity required by modern
machine learning and data analytics applications, or by large-scale 
high-fidelity scientific simulations, which typically offload compute 
intensive workloads to the attached GPUs for acceleration. As an example, 
in the latest release (Nov 2020) of the top500 list~\cite{top500}, 6 out 
of the top 10 HPC systems are equipped with powerful high-end GPU devices 
to deliver high-peak overall system performance. For many of these systems, 
each computing node is equipped with multiple GPU devices. A typical example 
is the Summit supercomputer, in which each compute node has $2$ IBM 22-core 
Power9 CPUs and $6$ NVIDIA Tesla V100 GPUs. By leveraging these GPU-powered 
HPC systems, it can run advanced, large-scale applications more efficiently 
and is touted to deliver 200 peta-flops peak performance. 

However, how to efficiently utilize these high-power GPU resources remains 
an open research problem in many contexts. While certain heavy, performance-critical 
workloads may require dedicated GPUs and are able to fully saturate these high-end 
devices, many others do not utilize these resources continuously to their 
maximum capacities~\cite{WLZJ2017}. Per a discussion with scientists from 
Los Alamos National Laboratory, a single scientific workload typically only 
uses $\sim30\%$ of GPU resources, leaving the majority of computing resources 
under-utilized and wasted. This trend also applies to machine learning 
workloads in data centers~\cite{XBRS2018}. The problem is exacerbated
because new generations of GPUs are very expensive,
both in terms of cost and power consumption. A high-end NVIDIA 
GPU device could cost as much as 2 to $5\times$ that of a high-end Intel 
Xeon CPU, and, in data centers, a GPU VM~(virtual machine) instance could be 
$10\times$ more expensive than a regular one. These practical observations 
demonstrate the necessity of efficient mechanisms to share GPUs among different 
workloads~\cite{GGHS2009,DPSM2010,GSTT2011,SCSL2012,TGCR2014,WLZJ2017,XBRS2018}, 
thereby increasing utilization, saving on energy consumption, and improving the 
cost-efficiency as well as throughput for these systems.

NVIDIA has been improving 
the support for GPU sharing in recent generations of GPUs (e.g. Volta) via the 
Multi-Process Service (MPS), which can run kernels from \emph{cooperative} 
processes (e.g. processes from the same MPI job) concurrently on a single GPU 
device. However, efficient sharing of \emph{multiple GPUs} (that reside inside a 
single, high-performance node) among \emph{uncooperative and independent} workloads 
remains an unsolved problem and poses the following challenges:

\begin{enumerate}
 \item While MPS can facilitate co-execution of kernels from different 
 processes on a single GPU device, it is mainly designed for cooperative 
 multi-process applications, e.g. MPI jobs. This kind of setting
 relies on programmers' knowledge to
 statically schedule kernel launches from different MPI ranks in a cooperative 
 way to avoid potential device overloading. Such a scheme is untenable when
 co-executing kernels from independent workloads, which have no knowledge 
 about other concurrent requests that are pending in terms of  GPU resources 
 (memory and cores), and the global picture of resource availability across 
 multiple devices. In particular, it has no support to ensure memory safety 
 of an executing kernel. If the available memory capacity is exceeded, the 
 process requesting the memory will crash due to the ``out of memory'' (OOM) 
 error, which could be disastrous for long-running applications. It also has 
 no knowledge of whether a given GPU's compute resources are saturated;
 over-saturation will slow down execution of individual applications, 
 whereas under-saturation leads to poor utilization. 

 \item MPS cannot schedule kernel launches across different GPU devices,
 and in current programming models, such as CUDA, programmers have to 
 explicitly and statically map kernel launches to expected executing 
 devices. They assume that an application has a dedicated access to the device(s),
 and have no knowledge about workloads on each device. Therefore, when multiple 
 independent workloads execute concurrently, some of the GPU devices 
 could be extremely overloaded while others may be idle. The imbalance between 
 GPU devices could adversely affect the execution efficiency (kernel execution 
 time as well as throughput). 
\end{enumerate}



The above limitations imply a need for system-level mechanisms to coordinate 
the execution of kernels {\it from independent and uncooperative applications}
on a set of GPU devices, therefore increasing the resource utilization, 
saving on energy consumption and improving the cost-efficiency, but incurring 
negligible performance interference for individual workloads. Our approach to
this challenging problem is a compiler-guided scheduling framework which uniformly
manages GPU resources for executing applications. Our solution is fully
automated without any manual effort or changes to application source 
code. It leverages the compiler coupled with a runtime system to construct 
GPU tasks, which are basic scheduling units in our scheduling framework.
Briefly, a GPU task contains one or more kernel launches, as well as other 
related GPU operations, e.g. GPU memory allocations and initialization, that 
are required to execute the underlying kernel(s) appropriately. A GPU task is generated through both static
analysis by the compiler and a lazy runtime by bundling together all the kernels that share underlying memory or exhibit
memory dependencies. Obviously, each 
GPU task contains a complete set of GPU operations required to finish a 
GPU computation, thus it can be scheduled and executed on any GPU device 
without breaking its correctness. For each GPU task, a probe is statically 
instrumented into its host-side code to gather and convey the task's resource 
requirements (such as memory footprints and number of SMs) to a user-level 
scheduler at runtime before the task is executed. Different scheduling policies 
can be implemented and deployed along with the proposed framework to target 
different computing environments. In this paper, we evaluate the technique's 
advantages by implementing a throughput-oriented scheduler for batch jobs
(such as ML training, data classification/analytics, linear algebra, etc.),
which are very important and popular in modern HPC/clouds~\cite{diversity}.
For such batch jobs, improving the system throughput is the first priority, 
and other characteristics such as fairness and QoS are not essential. The scheduler 
places tasks onto devices based on their resource requirements and the 
availability of each device. Such a 
scheme dynamically balances the workloads among GPU devices, and ensures
memory-safe executions in an environment shared by many independent workloads 
with almost no performance degradation. Therefore it significantly 
improves the system throughput as compared to state-of-the-art. To 
the best of our knowledge, this is the first work that aims
for a fully automated, efficient sharing of a multi-GPU system
among applications from different users.

In particular, this work makes the following contributions:
\begin{enumerate}
    \item We propose \emph{a GPU scheduling framework} to uniformly and 
    transparently manage GPU resources for applications. Utilizing this 
    framework, independent and uncooperative applications from different 
    users can simultaneously execute on a set of shared devices without 
    incurring  performance degradation caused by potential resource 
    contentions. The framework smartly places GPU tasks from different 
    applications on appropriate devices for best performance. 
    
    \item We devise a compiler pass, coupled with a lazy runtime, to construct 
    GPU tasks and insert probes that gather each of their resource requirements 
    (e.g. global memory and SMs). The probes convey this information to the 
    scheduler at runtime. Such a compiler-based solution allows one to construct 
    GPU tasks that can be dynamically bound to any GPU device at run time, yields 
    a precise analysis of resource requirements for each task, and fully automates 
    scheduling for these types of GPU applications.
    
    
    \item We implemented a prototype of the proposed framework on top of 
    NVIDIA GPUs, the CUDA library and the LLVM framework. Along with it 
    we also designed an efficient and fast throughput-oriented scheduling 
    algorithm to demonstrate the advantages of the proposed framework. 
    The algorithm quickly determines where a task should be executed 
    based on tasks' resource requirements and the current device workloads. 
    It guarantees the task to be executed efficiently and safely (without 
    OOM errors) by not overloading any device. We evaluate it with the 
    Rodinia benchmark suite and the Darknet neural network framework. The 
    results show that such a compiler-based, fully-automated solution outperforms 
    state-of-the-art frameworks due to its ability of leveraging applications' 
    knowledge about resource requirements, such that the scheduler can avoid 
    OOM errors and balance workloads among devices. On average, it can help 
    to improve the throughput of the system by over 2$\times$. 
\end{enumerate}

The rest of paper is organized as follows: Section~\ref{sec:background} introduces
the background and the motivation of this work. Section~\ref{sec:design} and 
Section~\ref{sec:impl} presents the detailed design and prototype of the proposed 
framework. Evaluation results are presented in Section~\ref{sec:eval}.
Finally, Section~\ref{sec:related} discusses the related state-of-the-art,
and Section~\ref{sec:concl} concludes our contributions and findings.

\section{Background}\label{sec:background}
In this section, we provide the necessary background
and challenges that motivate the design
of our compiler-guided scheduler. We follow the terminology 
used by NVIDIA GPUs and toolkit. GPUs from other vendors 
and programming models share a similar 
design paradigm. 

\subsection{GPU Architecture and Execution Model}

Modern GPU devices have a massive number of simple cores, 
which are grouped into multiple streaming multiprocessors (SMs).
For example, the NVIDIA P100 GPU has $56$ SMs, with each SM 
consisting of 64 single-precision CUDA cores, and thus 3584 CUDA cores in total. 
When a kernel is launched, a hardware scheduler 
is in charge of dispatching thread blocks (TBs) of the kernel to 
SMs, one at a time, in a round-robin fashion. TBs 
are basic scheduling units. 
They are independent of each other, and can be executed on 
different SMs in parallel. The hardware scheduler repeatedly dispatches 
a TB to an SM if the SM has sufficient hardware resources, until one of 
the required resources (e.g. registers or the maximum 
number of threads) reaches the SM limit. If the total GPU resources 
are not enough to accommodate the execution of all TBs in a kernel, the 
remaining TBs will wait for previous TBs to finish and release resources. 
Once dispatched, the threads within the TB are grouped into warps, where 
each warp contains $32$ threads. The warp is the basic scheduling unit 
inside an SM.
Once a kernel starts execution, it blocks all other kernels until all of 
its TBs are scheduled.

In older generations of GPUs, a device could be occupied by only one process 
at a time. Thus, if the process did not consume the maximum 
capacity of the device, the unused resources were wasted due to 
a lack of execution support for multiple processes. To mitigate this 
issue, MPS was introduced to allow kernels from different processes 
to simultaneously execute on the same device. However, MPS is mainly designed 
for co-operative multi-process CUDA applications, e.g. MPI jobs. 
These jobs rely on the programmers' knowledge to 
statically schedule GPU kernels from different MPI ranks in a 
cooperative way, which is an extremely daunting process for 
programmers. Such a process is error-prone and can lead
to memory errors and application crashes.
For today's environments, MPS lacks a couple of 
key features. 
First, it does not manage co-executing applications across
multiple GPU systems. 
MPS is only capable of managing and scheduling SMs for a 
single GPU among CUDA kernels. 
As a consequence, given a saturated GPU, MPS can only queue
kernels, resulting in performance penalties. It is unable to
move the queued kernels to other under-utilized devices.
(Part of this is also due to the programming 
model, which does not allow a dynamic binding of a kernel to 
a device, as explained in the next section.) Second,
it does not manage memory resources, which is left
to programmers. Thus, when multiple, independent (and independently
programmed) jobs are co-executing, applications will crash
if the memory capacity is exceeded.

\subsection{Sharing in a Multi-GPU System}

\begin{figure*}[ht!]
    \centering
    \subfloat[The System is Dedicated to App1]{
    \includegraphics[height=0.12\textwidth]{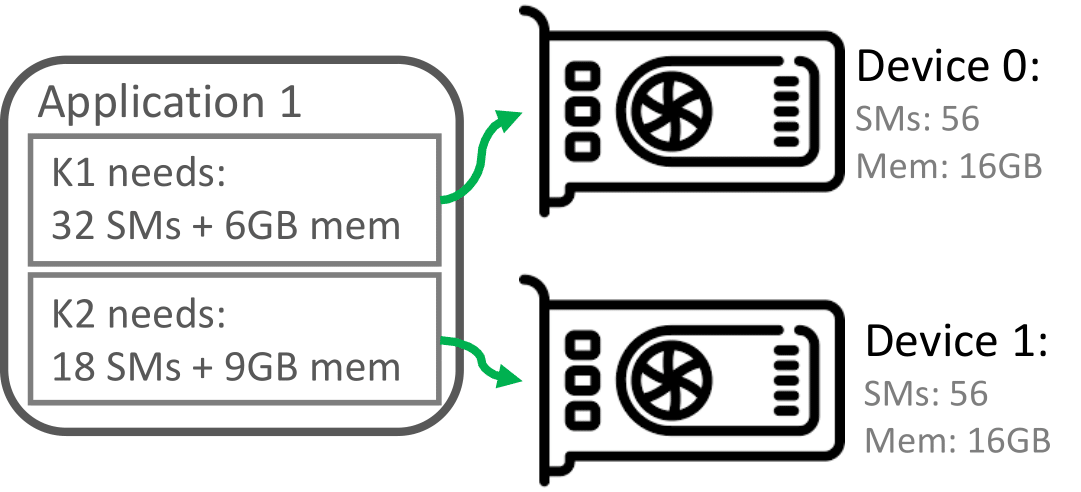}
    }
    \hfill
    \subfloat[The System is Dedicated to App2]{
    \includegraphics[height=0.12\textwidth]{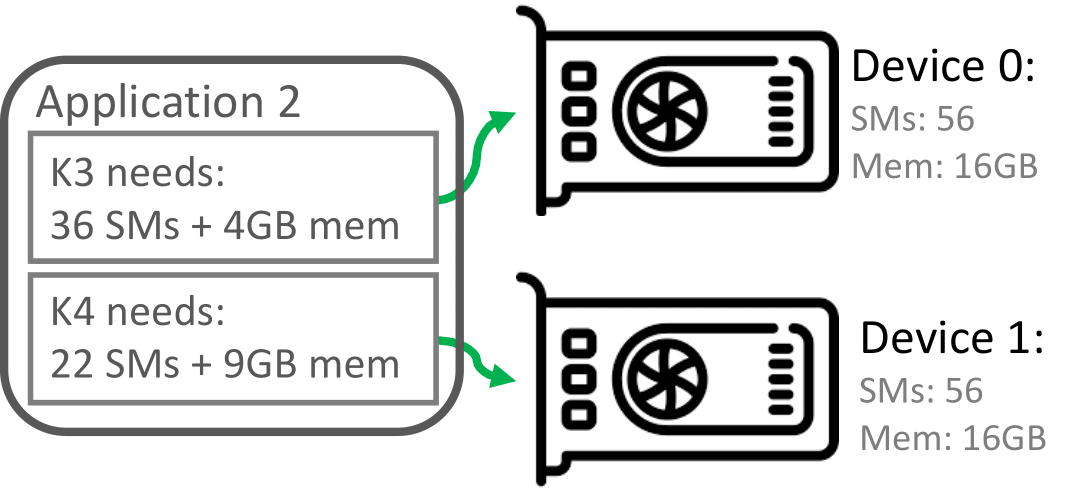}
    }
    \hfill
    \subfloat[The System is Shared by App1 \& App2]{
    \includegraphics[height=0.12\textwidth]{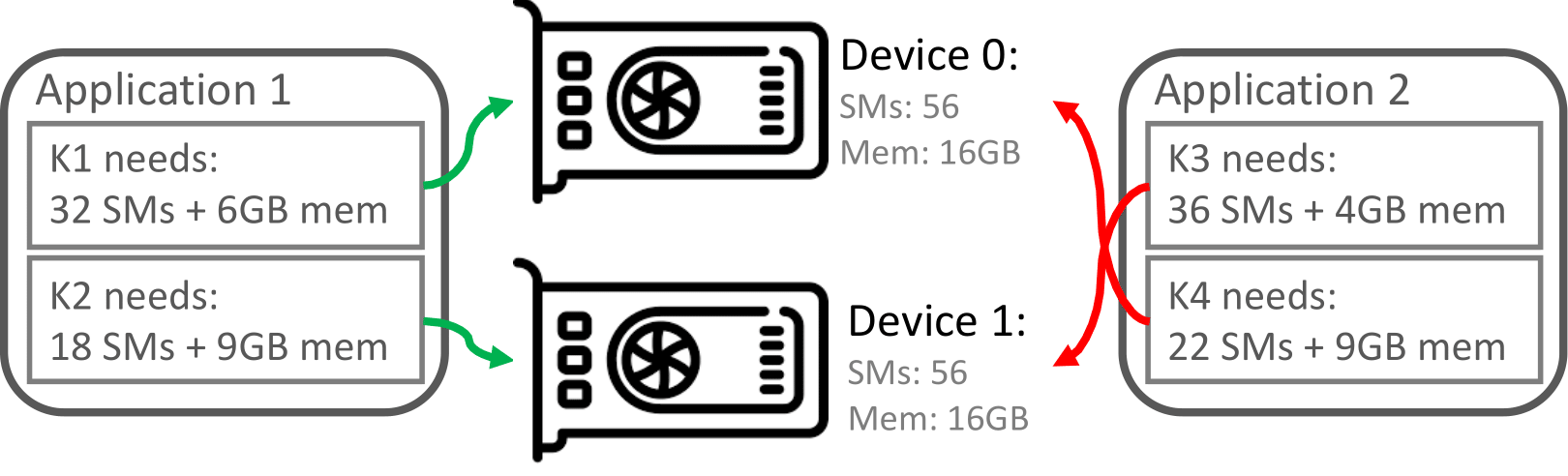}
    }
    \caption{Kernel Mapping}
    \vspace{-.0in}
    \label{fig:taskmap}
\end{figure*}

In a multi-GPU system (a single compute node equipped with multiple GPU devices), 
applications can distribute their CUDA kernels on different devices to execute
them in parallel, thereby maximizing their performance. However, such
systems currently 
rely on programmers to explicitly designate an execution device for each kernel  
launch and its related CUDA operations through the {\tt cudaSetDevice} API. 
If there is no such call in the application, the CUDA runtime will bind 
every CUDA operation to $device 0$ by default. Obviously, this static scheduling 
solution requires significant 
programming effort to explicitly and carefully designate a device for each 
CUDA operation, ensuring that the device has enough resources to execute assigned 
kernels. While it may be somewhat viable (but tedious) on an
application-by-application basis,
such a static binding is not feasible for applications from different users due to the absence of the 
knowledge about the resource requirements and dynamic concurrency of the executing applications. 

Figure~\ref{fig:taskmap} illustrates the issue with an example on a 
2-GPU system. Each GPU has $56$ SMs and $16$GB DRAM. 
It assumes there are $2$ applications, with each having 2 CUDA 
kernels that can be executed in parallel, and each kernel needing 
different GPU resources. If the system is dedicated to each of these 
applications, it is easy to achieve good performance by simply 
mapping kernel $k1$ to device $0$ and kernel $k2$ to device $1$ 
for $application1$, and mapping kernel $k3$ to device $0$ and 
kernel $k4$ to device $1$ for $application2$. 
By closely examining the resource requirements 
for each kernel, one can see that it is possible to share the system 
between these two applications without performance degradation, because 
their total resource requirements are within the system capacity. 
However, the previous statically determined schedule (mapping) will
not work in this shared scenario, because the total 
SM requirements of $k1$ and $k3$, and the total memory requirements 
of $k2$ and $k4$ exceed the capacity of a device. While the overload 
of SM resources could cause performance interference and
degradation, the overload of 
memory will lead to application failures due to the OOM error. 
For this example, the solution is to co-locate $k1$ and $k4$ 
on a device, and co-locate $k2$ and $k3$ on another device. However, it is 
impossible to make such a decision statically, and a dynamic, runtime solution 
is proposed in this work. The proposed method manages GPU 
resources uniformly and allocates them at each kernel launch per request. 
Then the kernel will be scheduled on an appropriate device based on its resource 
requirements and status of each device to ensure the memory safety and minimize 
the performance interference among workloads.

\section{The Proposed Framework}\label{sec:design}

\begin{figure}[!tp]
    \centering
    \includegraphics[width=0.9\columnwidth]{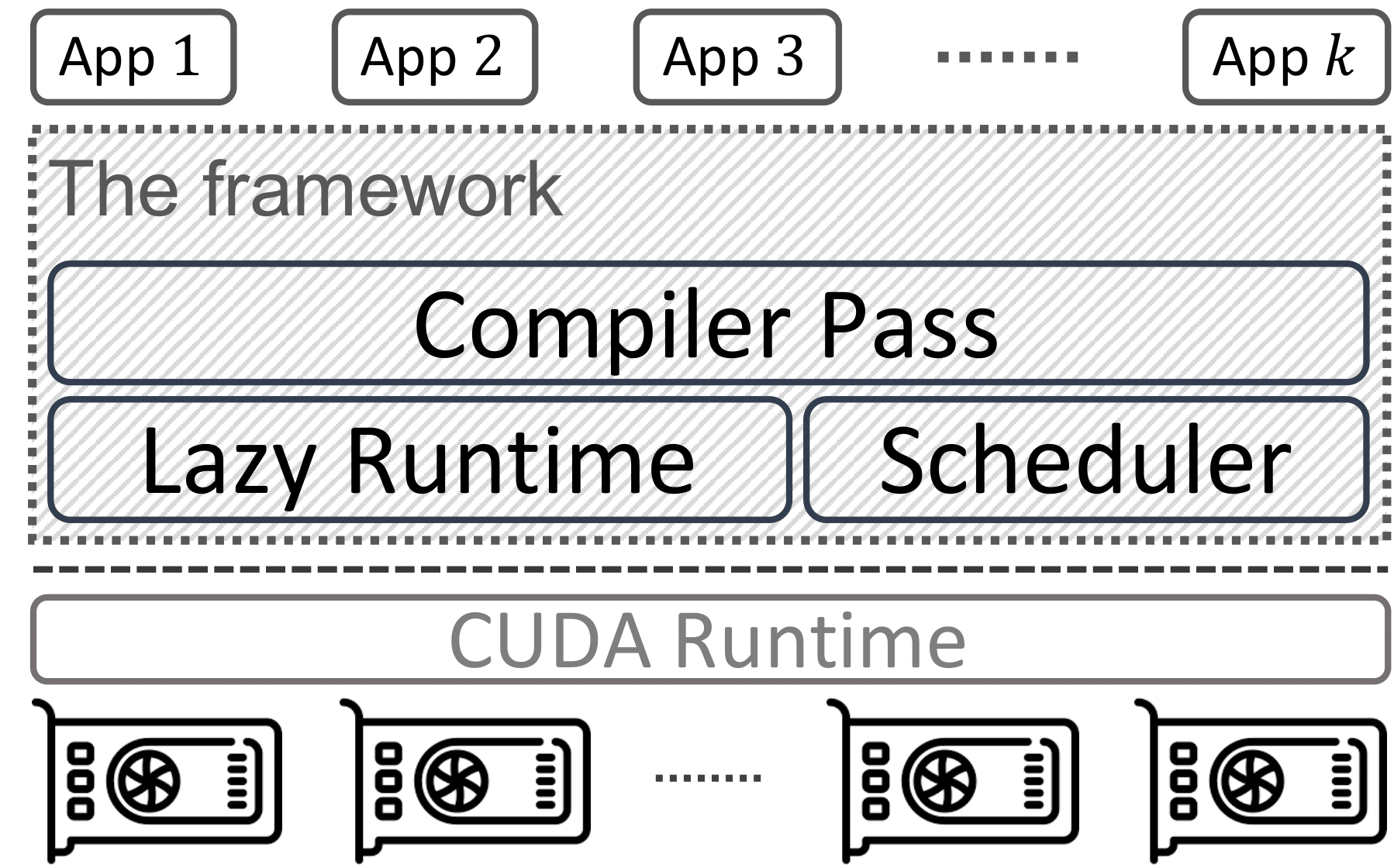}
    \caption{Overview}
    \vspace{-.15in}
    \label{fig:overview}
\end{figure}



To achieve the above goals, the proposed framework consists of 
three main components: a compiler pass, lazy runtime, and scheduler
(see Figure~\ref{fig:overview}). The compiler pass, coupled with the 
lazy runtime, constructs GPU tasks and instruments applications with 
probes, one per GPU task. At runtime, the probes will convey the 
resource requirements of GPU tasks to the scheduler before they 
are executed, and the scheduler then assigns GPU tasks to appropriate 
devices by interpreting their probes and tracking devices' statuses. 

\subsection{GPU Tasks}\label{sec:cudatask}

\begin{figure}[!tbp]
\begin{lstlisting}[language=C, style=c]
// VecAdd is a CUDA kernel executed on GPU
__global__ void VecAdd(int *A, int *B, int *C) {
  int i = blockIdx.x * blockDim.x + threadIdx.x;
  C[i] = A[i] + B[i];
}

// main is sequential code running on CPU
int main(int argc, char **argv) {
  int A[N], B[N], C[N], *dA, *dB *dC;
  
  // initialize the vectors
  for (int i = 0; i < N; i++) {
      A[i] = cos(i);
      B[i] = sin(i);
      C[i] = 0;
  }
  task\_begin(N*3, 128, N/128); // the prob instrumented
  // allocate device memory
  cudaMalloc(&dA, N); // an input vector
  cudaMalloc(&dB, N); // an input vector
  cudaMalloc(&dC, N); // for storing result
  
  // initialize the device memory
  cudaMemcpy(dA, A, N, cudaMemcpyHostToDevice);
  cudaMemcpy(dB, B, N, cudaMemcpyHostToDevice);

  // launch the kernel on device
  dim3 T(128), B(N/128);
  VAdd<<<B, T>>>(d_A, d_B, d_C);
  
  // retrieve the result
  cudaMemcpy(C, dC, N, cudaMemcpyDeviceToHost);

  cudaFree(dA); 
  cudaFree(dB); 
  cudaFree(dC);
}
\end{lstlisting}
\vspace{-.1in}
\caption{An example GPU task, which consists of a kernel launch and related 
GPU memory operations.}
\label{lst:task}
\vspace{-.1in}
\end{figure}

The ``GPU task'' is the basic scheduling unit in the framework. 
It is a collection of GPU operations that are a preamble to a 
GPU kernel (such as memory allocation, data transfer, etc.) 
along with the kernel launch itself. 
While the core operation of a GPU task is a kernel launch, it is 
necessary to do several other GPU operations to set up the execution 
context to facilitate the correct execution of the kernel. Typically, 
it includes allocating space on the target device (e.g., \codeword{cudaMalloc}), 
and initializing these spaces with required data (e.g., \codeword{cudaMemcpy} 
or \codeword{cudaMemset}). Finally, after the execution of the kernel, 
the results of the job need to be saved, and the allocated memory must 
be freed (e.g., \codeword{cudaFree})~\footnote{It is obviously
invalid to execute {\tt cudaMalloc} and {\tt cudaMemcpy} operations on one 
device, but execute the kernel and the rest of the operations on another.}.
All of these related GPU operations should be issued to the same device
and form a GPU task. An example of GPU task is 
shown in Figure~\ref{lst:task}, in which the code from line 19 $\sim$ 
36 form a GPU task for adding two vectors and getting its results. The 
task consists of a kernel launch (line 22), as well as related memory 
operations (line 19 $\sim$ 25 and line 32 $\sim$ 36) for preparing the 
memory and data required by the kernel and releasing the occupied resources 
upon existence.
 
\subsubsection{Task Construction}
The framework leverages a compiler pass, coupled with the lazy 
runtime, to construct GPU tasks and gather their resource requirements. 
The compiler pass works on LLVM IR of applications. It builds GPU 
tasks around kernel launches as indicated by calls to specific runtime 
APIs (e.g., {\tt \_cudaPushCallConfiguration}). For each kernel launch, 
the thread information (grid and block) is first retrieved 
by examining related parameters of the call. The host 
stub code for the kernel launch is then identified after 
the call to, e.g., {\tt \_cudaPushCallConfiguration}, by following 
the control flow of the program, from which the memory objects 
(in the form of pointer variables) accessed by the kernel are 
extracted. The compiler pass uses these memory objects to 
identify all other related GPU operations (e.g {\tt cudaMalloc}, 
{\tt cudaMemcpy}, {\tt cudaFree}, etc.) based on def-use chains 
of these LLVM IR values, and then constructs the GPU task 
(GPUUnitTask in \ref{alg:buildtask} ) based on 
either dominator information (for {\tt cudaMalloc} and host-to-device 
{\tt cudaMemcpy} operations) or post-dominator information (for 
{\tt cudaFree} and device-to-host {\tt cudaMemcpy} operations)
as regarding to the kernel launch.
If several GPU tasks 
share a set of memory objects, they will be merged into a larger task 
(GPUTask in \ref{alg:buildtask})
based on the observation that it is inappropriate to schedule them on 
different devices due to the memory dependency and the costs of moving 
data among devices. A typical example is a process executing two 
successive GPU kernels, $k1$ and $k2$, where the output of $k1$ (say, 
array \codeword{C}) is an input to $k2$. If $k1$ and $k2$ are scheduled
onto two different devices, the data for \codeword{C} needs to be copied to the 
device running $k2$. To avoid the cost of such data movement, the 
framework schedules these two kernel launches on the same 
device by packing them into one GPU task. Algorithm~\ref{alg:buildtask} 
outlines this approach for building GPU tasks. 
Finally, the memory and computing resource 
requirements of the task can be analyzed by examining every memory 
allocation operations ({\tt cudaMalloc}) and kernel launch operations 
(e.g., {\tt \_cudaPushCallConfiguration}) inside the GPU task. It is 
worth noting that all of the analyzed information is in the form of 
symbols, and a probe is inserted at a program point which post-dominates 
all of these symbol definitions and dominates all GPU operations in the task. 
The probe will interpret these symbols at runtime to get actual resource 
requirements for each GPU task, and convey them to the underlying 
user-level scheduler.


\subsubsection{Lazy Runtime}
Many applications encapsulate kernel launches and other GPU operations 
in separate functions, e.g. allocating GPU memory in {\tt init()} 
and launching kernels in {\tt execute()}. Static analysis is unable to establish
such def-use chains 
and domination relationships inter-procedurally among GPU operations~\footnote{These two 
compiler analyses are performed intra-procedurally in our framework}.
To mitigate this issue, an inlining pass is first leveraged.
If it cannot address the problem,
the compiler will defer the bindings of the memory operations to a task through the lazy runtime, which works as follows.

The statically unbound operations are marked for lazy binding by the compiler. 
This enables the lazy runtime to record all of GPU 
operations and delay their bindings (executions) until a kernel 
launch. For example, a call to {\tt cudaMalloc} will be replaced 
with {\tt lazyMalloc}, which will simply assign a pseudo 
address for representing the memory object to be allocated, instead 
of performing the actual allocation. Thus, the subsequent CUDA 
operations on the memory object will see the pseudo address (and in fact
all those CUDA operations are replaced with corresponding lazy runtime
operations, as well). For each memory object, a queue is maintained to record GPU 
operations applied on it (e.g., [{\tt cudaMalloc}, {\tt cudaMemcpy}]) in 
execution order. Just before every kernel launch operation (e.g., {\tt \_\_cudaPushCallConfiguration}), 
a specific lazy runtime API {\tt kernelLaunchPrepare} is inserted. It will 
interpret the memory objects needed by the kernel, replay the recorded 
GPU operations for each of them, and replace their pseudo addresses with 
the real ones to ensure the kernel can be executed successfully. It 
also collects the resource requirements of the kernel launch by associating (or binding) them to 
the CUDA task being launched and conveys 
them to the scheduler.  
Such an approach, coupled with the above static program analysis, binds 
full resource needs to a kernel, thereby converting it into a device-independent 
entity for the scheduler. The scheduler can then assign the task dynamically 
to a device and allocate the required resources recorded in the probes. 

\subsubsection{On-device Dynamic Allocation}
In addition to global memory allocations, dynamic memory allocation from 
inside a kernel also need to be considered. While it could be difficult 
to get accurate memory resources that will be allocated inside a kernel, 
it is easy to get the upper bound based on current GPU runtime and 
architecture design. For example, the on-device heap size defaults 
to 8MB for the NVIDIA devices we tested. Applications can increase this 
limit by adjusting the {\tt cudaLimitMallocHeapSize} via a call to 
{\tt cudaDeviceSetLimit}; and this call must be placed before launching the kernel. 
Thus, the maximum heap memory size used by dynamic memory allocations 
inside a GPU is either statically bound to a CUDA task or dynamically intercepted and bound
by the lazy run time by analyzing the call to 
{\tt cudaDeviceSetLimit}.

\begin{algorithm}[tbp]
\caption{The pseudo code of constructing GPU tasks using static program analysis}
\label{alg:buildtask}
\begin{algorithmic}[]
\Function{buildGPUTasks}{ }
    \State vector$\langle$GPUTask$\rangle$ $Tasks$
    \State vector$\langle$GPUUnitTask$\rangle$ $UnitTasks$
    \For{each kernel launch $l$}
    \State $memObjs \gets \Call{getMemArgs}{l}$
    \State $allocs \gets \Call{getAllocOps}{l}$
    \State $blocks \gets \Call{getGridDims}{l}$
    \State $threads \gets \Call{getBlockDims}{l}$
    \State $UnitTasks$.push($blocks$, $threads$, $allocs$, $l$)
    \EndFor
    
    \For{each unvisited unit task $u1$ in $UnitTasks$}
    \State set$\langle$CUDAUnitTask$\rangle$ $Union$;
    \State $visited$[$u1$] $\gets$ true
    \For{each unvisited unit task $u2$ in $UnitTasks$}
    \If{$u1.memobjs \cap u2.memobjs \neq \emptyset$}
    \State $Union$.insert($u1$, $u2$)
    \State $visited$[$u2$] $\gets$ true
    \EndIf
    \EndFor
    \If{$Union$.size == 0}
    \State $Tasks$.push($u1$)
    \Else
    \State $Tasks$.push(merge($Union$))
    \EndIf
    \EndFor \\
    \ \ \ \ \ \ \Return $Tasks$
\EndFunction
\end{algorithmic}
\end{algorithm}

\subsection{The Scheduler}
A user-level scheduler is designed to place GPU tasks to appropriate devices 
based on their resource requirements (such as memory, CUDA cores, shared 
memory and execution time of a kernel). The scheduler exposes a simple API, 
{\tt task\_begin}, which will be automatically instrumented by the compiler 
at the beginning of each GPU task to deliver their resource requirements 
to the scheduler. Based on their resource requirements, the scheduler 
will find a appropriate device that meets the requirements based on a 
scheduling policy. While different scheduling policies can be implemented 
for targeting different computing environments, in this paper, we implemented 
a throughput-oriented scheduling policy for batch jobs to demonstrate the 
advantages of the proposed scheduling framework. We choose throughput-oriented 
scheduling policy because in modern systems it demonstrates a dominant usage
pattern of GPU sharing. A throughput oriented batch scheduler is 
used in many scenarios in modern HPC/clouds~\cite{diversity, borg}, for workloads such as 
ML training, data classification/analysis, linear algebra etc. where large jobs must complete
as a batch as fast as possible. For these 
batch workloads, the fairness and QoS are not important but the throughput is the important
systems goal. Due to this motivation, we demonstrate the effectiveness of sharing through 
the throughput improvement of the system.



\begin{algorithm}[!tpb]
\caption{The pseudo code to select a GPU for a process,
emulating the way hardware tracks SM usage}
\label{alg:schedule}
\begin{algorithmic}[]
\Function{sched}{$task, GPUs$}
    \State $TargetG \gets None$
    \For{$G$ in $GPUs$}
      \State $TBs \gets task.ThreadBlocks$
      \If {$task.MemReq$ $>$ $G.FreeMem$}
        \State continue;
      \EndIf
      \While {$TBs > 0$}
        \State $availSM \gets G.\Call{getNextSM}{task}$
        \If{!$availSM$}
          \State break
        \EndIf
        \State $availSM.\Call{add}{TB}$
        \State $TBs--$
      \EndWhile
      \If{$TBs == 0$}
        \State $G.\Call{commitSMChanges} $
        \State $TargetG \gets G$
        \State break
      \EndIf
    \EndFor
    \State \Return $TargetG$
\EndFunction
\end{algorithmic}
\end{algorithm}

The scheduling policy makes the decision based on a vector of metrics 
including the availability of global memory, as well as SMs. The 
multi-resource oriented scheduling problem is NP-hard. In this paper, 
we look at two scheduling algorithms that are tailored specifically to 
the problem at hand. Algorithm~\ref{alg:schedule} emulates
hardware's round-robin approach for placing a task's
thread blocks across a GPU's SMs.
It tracks exactly how many thread blocks and warps on 
each SM are available (taking into account the device's max thread 
blocks and warps per SM). It also ensures
that the memory required by a task is available on the
selected GPU. Both memory and compute are hard constraints in this
algorithm. In contrast, Alg.~\ref{alg:schedule2} is simpler.
It treats memory as a hard constraint, but it treats compute as a soft 
constraint (because it can impact
performance but will not lead to a crash). First it checks if the
memory requirement of incoming task can be met on a GPU device and comes
up with a list of all devices that satisfy the memory requirements. Next,
from this list of devices, it picks the device with the least load in terms
of number of warps currently scheduled on it. In other words, it
simply tracks the total number of active warps on a GPU (not
at a granular, SM level), and picks the GPU 
with the least load in terms of the total number of warps scheduled carrying out a fast scheduling decision.
It is not as accurate as 
Alg.~\ref{alg:schedule}, but it can make quicker decisions and thus can
take advantage of dynamic opportunities (such as fast
task completions) 
that might have arisen during the scheduling decision window. Once 
a GPU is selected for a task, both the available memory
and warp capacity of the GPU are updated. Both scheduling algorithms 
are designed to be very simple
to minimize the runtime overheads and to keep them dynamically reactive to short GPU jobs.

\section{Prototype Implementation}~\label{sec:impl}
We implemented a prototype of the proposed scheduling framework 
based on NVIDIA GPUs, CUDA-10.2 and LLVM-9.0. The compiler 
pass works on the LLVM IR representations of applications. 
For convenience, we will begin referring to the proposed 
framework as \shortname (for ``multi-GPU bearer''). 

As mentioned before, {\tt task\_begin} is instrumented at the beginning 
of each GPU task to convey its resource requirements to the scheduler. 
In response, the scheduler returns the ID of the device 
where the task will be executed. And then {\tt task\_begin} calls 
{\tt cudaSetDevice} to map the task to the target device. For each 
GPU device, the MPS is enabled such that the kernels from different 
processes can run on the same device. Once a GPU task completes and 
associated resources are released, \shortname will attempt to pack 
more processes in the newly available space. All communication between
processes and the scheduler is implemented over shared memory.


For comparison, we also implemented two other scheduling policies,
including {\it single-assignment (\sa) scheduling} and the {\it 
Core-to-GPU (\cg) scheduling}, to mimic current practical strategies 
of sharing multiple GPUs among independent workloads. Both of them 
make scheduling decisions without the knowledge of tasks' requirements. 
they provides a good baseline for analyzing the proposed \shortname 
framework to demonstrate the advantages of leveraging applications' 
knowledge. In section \ref{sec:neural}, we also provide comparisons 
against~\cite{RSNV2018} for resource-heavy machine learning workloads. 
We don't compare our work to other schedulers, e.g., FLEP~\cite{WLZJ2017}, 
mainly because they do not handle the multi-GPU case, and are designed 
for QoS sensitive workloads sharing a single GPU device. And also because
they are not open-sourced such that we can port them to our framework.

\sa shares the same scheduling strategy of utilizing multiple 
devices among independent workloads with the work in~\cite{YoJG2003}. 
It distributes workloads among GPUs at process-level granularity. 
When a CUDA application begins, \sa maps it to the first available 
GPU device. \sa ensures each application has dedicated access to 
the assigned device during its lifetime. Each device has no more 
than one job at a time (assuring memory safety), and no device sits 
idle once a request is made.

Considering that a device could be extremely underutilized in 
\sa, \cg is designed to allow more than one process 
to share a GPU device via NVIDIA MPS. Therefore, \cg could 
be more performant than \sa in terms of better system throughput 
and device utilization. However, \cg is unsafe because many 
workloads could be terminated unexpectedly due to memory outages. 
To mitigate this issue, \cg attempts to control the maximum number of 
jobs per GPU through a pre-determined \cg ratio. The ratio may be heuristically derived 
based on system configurations. For example, in a system with 12 CPU 
cores, 2 GPUs, and mildly memory-hungry jobs, each device might serve 
kernels from no more than 6 cores (with 1 process per core), producing 
a \cg ratio of 6:1. In our experiment, we examined multiple \cg ratios. 
At runtime, the \cg scheduler will visit the GPU task queue in a round 
robin manner and map the tasks to GPU devices until the ratio is met 
(in the above example, 6 tasks will be mapped per GPU device). Since 
the scheduler has no knowledge of the memory requirements of the tasks, 
such a mapping still stands the risk of ``out-of-memory'' errors and 
crashes (which we observed in some cases). When it does not crash, it 
leads to better performance than \sa, but our experiments show that
\shortname outperforms \cg in such cases as well.

\begin{algorithm} [!tbp]
\caption{The pseudo code to select a GPU for a process,
with memory safety and quick placement based on max available warps}
\label{alg:schedule2}
\begin{algorithmic}[]
\Function{sched}{$task, GPUs$}
  \State $TargetG \gets None$
  \State $MinWarps \gets$ 0
  \For{$G$ in $GPUs$}
    \If {$task.MemReq$ $<$ $G.FreeMem$}
      \If {$MinWarps$ $<$ $G.InUseWarps$}
        \State $MinWarps \gets G.InUseWarps$
        \State $TargetG \gets G$
      \EndIf
    \EndIf
  \EndFor
  \If{$TargetG$}
    \State $TargetG.\Call{addWarps}{task}$
  \EndIf
  \State \Return $TargetG$
\EndFunction
\end{algorithmic}
\end{algorithm}

\section{Evaluation}\label{sec:eval}

%
%

We evaluated \shortname with both a 2-GPU system running NVIDIA P100s 
and Intel Xeon E5-2670 with 128GB of RAM from Chameleon\footnote{an 
experimental test-bed for computer science funded by the NSF Future 
Cloud program}, and a 4-GPU system running NVIDIA V100s and Intel 
Xeon E5-2686 with 244GB of RAM from AWS (p3.8xlarge instance). Each 
P100 has 16GB of RAM and 3584 cores, and each V100 has 16GB of RAM 
and 5120 cores. Our evaluation seeks to answer the following questions:
\begin{itemize}
    \item[1)] What is the throughput improvement due to \shortname over \sa?
    \item[2)] What is the crash behavior of a memory-unsafe scheduler
    that attempts to pack incoming tasks onto GPUs, i.e. a non-compiler guided solution like the \cg scheduler? 
    \item[3)] What is the improvement in the average job turnaround
    time, which is defined as the interval between job completion time and its arrival time in the queue? 
    \item[4)] What is the negative effect of \shortname on an individual kernel's execution speed?
    \item[5)] How much better is \shortname over competing schemes such as ~\cite{RSNV2018} which use memory footprint as a resource constraint but do not use the warps or thread-blocks needed?
\end{itemize}





\subsection{Workloads}\label{sec:workloads}
We leveraged the CUDA benchmarks in the Rodinia suite
v3.1~\cite{rodinia09, rodinia10} for creating
the majority of our job mixes. We also have a subsection devoted
to neural network workloads, where we leveraged Darknet~\cite{darknet13} for ML workloads.
For Rodinia, We selected benchmarks and their problem sizes
that generate modest-to-large memory footprints, representative of
modern workloads for these multi-GPU systems. In summary, we found
7 unique benchmarks that fit this criteria, and we varied their
arguments to give us a bigger pool for creating mixes.
These are backprop (pattern recognition), srad-v1 and srad-v2
(image processing), lavaMD (molecular dynamics), needle
(bioinformatics), dwt2d (image/video compression), and bfs (graph).
We have 7 benchmark-argument combinations 
that generate 1 to 4 GB of footprint (using using all but
lavaMD), and 10 which generate over 4 GB (using all but bfs).
The largest memory footprint is $\sim 13$ GB for a lavaMD instance.

Our mixes favor larger workloads to mimic realistic, heavy GPU kernels.
We mark benchmarks with kernels that have over a 4GB memory requirement
as ``large''. Those between 1 and 4 GB are considered small. Nothing
below 1 GB is used in our Rodinia experiments. Our mixes
are a ratio of large:small jobs. We have four different mixes: 1:1,
2:1, 3:1, and 5:1. Every mix matches one of these ratios, but the jobs are
randomly chosen from their respective sets. We generated workloads of
16 jobs and 32 jobs, which typically last up to 5 and 10 minutes to mimic long running jobs, respectively.
Thus, in total we have 8 Rodinia workloads (Table~\ref{tab:workloads}). 

\begin{table}[!tbp]
    \centering
        \caption{Workloads}
    \begin{tabular}{|c|c|c|c|}
    \hline
         Workload & mix & Workload & mix  \\ \hline
         W1 & 16-job,1:1-mix & W2 & 16-job,2:1-mix \\\hline
         W3 & 16-job,3:1-mix & W4 & 16-job,5:1-mix \\\hline
         W5 & 32-job,1:1-mix & W6 & 32-job,2:1-mix \\\hline
         W7 & 32-job,3:1-mix & W8 & 32-job,5:1-mix \\\hline
    \end{tabular}
    \label{tab:workloads}
\end{table}

The workloads are assumed to be queued at the time that each experiment
begins, mimicking batch processing, a processing setup that would stress the scheduling system to the maximum since every job is ready (as opposed to arriving, for example, at predetermined or random times).
A pool of workers is responsible for processing the batch of jobs. Each worker dequeues a job, runs it, and then pulls another
until the work is complete.

The number of workers is determined by the scheduler and experiment.
The \sa scheduler always has a number of workers 
equal to the number of GPUs. The number of
workers for the \cg scheduler varies with the experiment.
The \cg scheduler is unstable due to memory safety violations
and crashes often. For a fair comparison,
we swept different worker pool sizes for the \cg scheduler and took the 
best performing runs that did not crash.
For the \shortname scheduler, we sized the worker pool statically.
As the number of workers increases, better packing is possible;
but this can also slow the overall system, as well.
For example, in a 2:1 (large:small) mix of 16 jobs on a
2xP100 system, \shortname takes the same amount of time to complete
the workload with 6 workers as it does with 16 workers; whereas
10 workers are about 10\% faster.
We found that a reasonable median across our experiments for
\shortname was 10 workers for the 2xP100s and 16 workers for the
4xV100s. Of course, determining the right number of workers (statically
or dynamically) is a problem in itself, but we leave this as future work.

Due to the prevalence of machine learning workloads and in particular
neural networks, we include a study on solely
these types of jobs. We instrumented an off-the-shelf learning framework, Darknet.
It provides several facilities, including
common training and classification tasks.
Its pre-trained models for image classification are competitive
with popular networks like ResNet-50 \cite{resnet-50} and VGG-16 \cite{vgg-16}
(in terms of top-1 and top-5 accuracy, GPU timing, and size); and as
a framework it is effective for creating other types of neural network
tasks (such as RNN text generation).
We ran 4 types of jobs: neural network training and prediction for image
classification, real-time object detection, and RNN-based text generation.
For prediction, we used the pre-trained
Darknet19 and Darknet53-448x448 architectures and weights
for the 1000-class ImageNet competition \cite{imagenet};
for training, we used the small architecture provided by Darknet for
CIFAR-10 \cite{cifar-10}; for real-time object detection, we used
the pre-trained yolov3-tiny architecture and weights on the provided
images; for RNN text generation, we used the pre-trained network
based on Shakespeare's complete works.

Note that these types of workloads are not a typical use case
when trying to gather traditional GPU metrics
(achieved occupancy, stall memory dependencies, warp execution efficiency, etc.).
NVIDIA tooling is geared towards standalone
kernels or towards MPI tasks from a single application.
{\it nvprof} usually relies on replaying kernels in order to profile them,
but this would affect results in co-executing environments.
(We tried the disable-replay option, as well, but the metrics were
dropped.)
Lastly, the metrics are stored on chip in buffers, which interferes
with the scheduler packing. We had to therefore resort to
macro-measures such as effects on the execution speeds of individual
kernels in a co-executing environment of \shortname vs.
the single execution environment such as \sa.


\newcommand{\shortnamesm}{{\bf MGBsm}\xspace}
\subsection{\shortname Scheduling Algorithms: Comparison}
\begin{figure}[!tbp]
    \centering
    \includegraphics[width=\columnwidth]{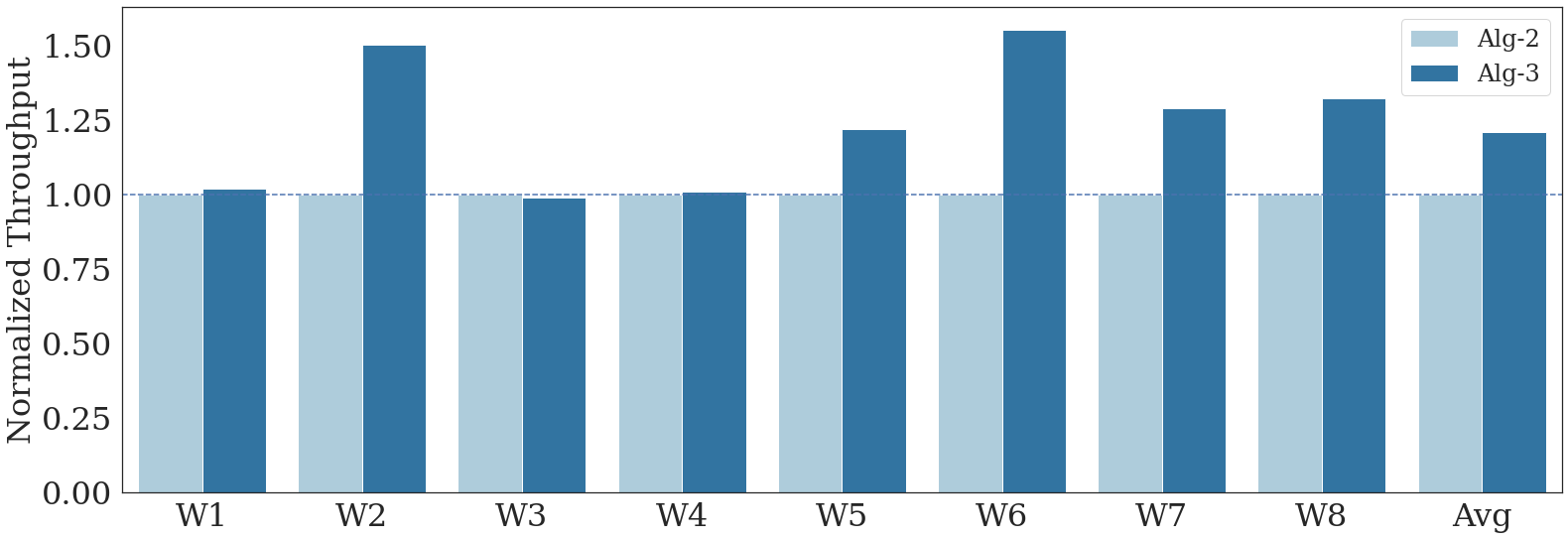}
    \caption{Throughput for Alg.~\ref{alg:schedule} and Alg.~\ref{alg:schedule2} 
            on a 4$\times$V100 system (normalized to Alg.~\ref{alg:schedule} 
            for easy comparison)}
    \vspace{-.1in}
    \label{figure:throughput-mgbsm}
\end{figure}

We compared different scheduling algorithms 
(Alg.~\ref{alg:schedule} and Alg.~\ref{alg:schedule2}) 
presented in section~\ref{sec:design}. Figure \ref{figure:throughput-mgbsm} 
shows their throughput evaluated with 8 workloads in a 4xV100 environment.
On average, the throughput for Alg.~\ref{alg:schedule2} is 1.21x higher.
We also scaled our experiments to 32 workers on 32-, 64-, and 128-job mixes,
and observed similar improvements.
Alg.~\ref{alg:schedule2} outperforms  Alg.~\ref{alg:schedule} mainly because of
the extra time jobs wait for a GPU under Alg.~\ref{alg:schedule}.
We observed a
30\% increase in Alg.~\ref{alg:schedule} in terms of job wait times over Alg.~\ref{alg:schedule2}. Alg.~\ref{alg:schedule} ensures there is sufficient
compute available before running each job, whereas Alg.~\ref{alg:schedule2}
schedules jobs optimistically and sooner, taking advantage of fast completing jobs, even when compute is stressed.
Alg.~\ref{alg:schedule2} seems to be relying partly on optimistically queuing jobs under MPS
and thus gaining throughput benefits.
In the rest of the section, we evaluated \shortname
with Alg.~\ref{alg:schedule2}, because it has better performance.

\subsection{Throughput}

\begin{figure*}
  \begin{subfigure}[b]{0.475\textwidth}
  \centering
      \includegraphics[width=\textwidth]{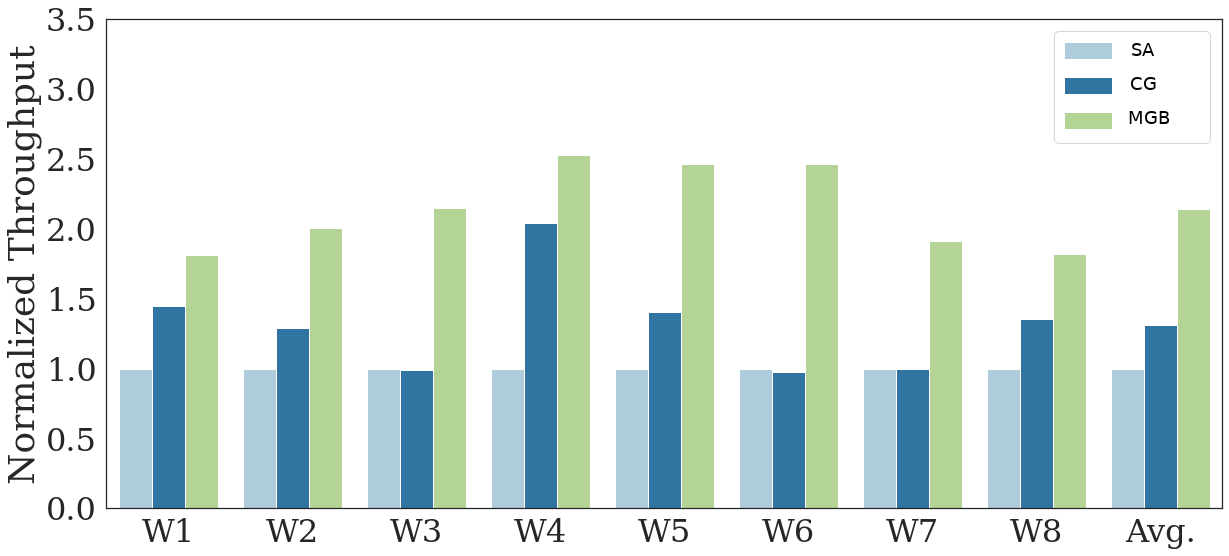}
    \caption{on a 2$\times$P100 system}
    \label{figure:throughput_p100}
  \end{subfigure}
  \qquad
  \begin{subfigure}[b]{0.475\textwidth}
      \centering
      \includegraphics[width=\textwidth]{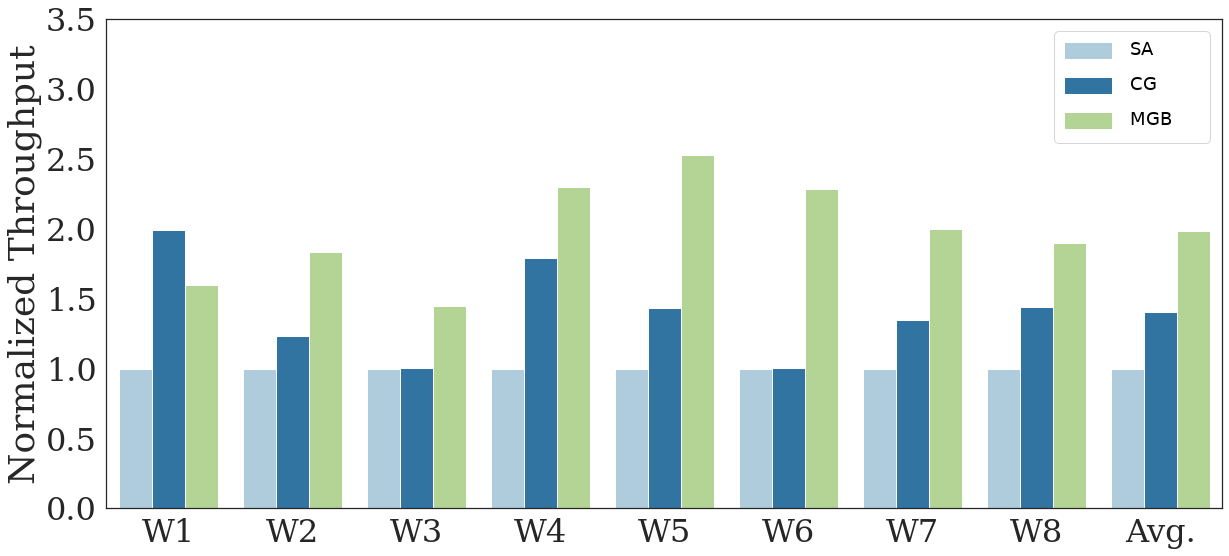}
      \caption{on a 4$\times$V100 system }
      \label{figure:throughput_v100}
  \end{subfigure}
  \caption{Throughput for \sa, \cg, and \shortname (normalized to \sa)}
  \vspace{-.1in}
  \label{figure:throughput}
\end{figure*}

Figure~\ref{figure:throughput} compares the throughput of each 
scheduler on two evaluated platforms. The throughput is normalized 
to \sa. Compared to \sa, 
\shortname improved system throughput by $1.8\sim 2.5 \times$ (on 
average $2.2\times$) on P100s and $1.4\sim 2.5 \times$ (on average 
$2\times$) on V100s. This is mainly because \shortname allows multiple 
kernels from different processes to be concurrently executed on the 
same device. Although the \cg scheduler also allows co-execution of kernels 
from different processes,
\shortname improved throughput by an average of 64\%
on P100s and 41\%
on V100s compared to \cg.
This is mainly because \cg has no knowledge about the memory or 
SM requirements of workloads; therefore it could overload GPU devices
and cause some jobs to crash due to memory safety violations
(see Table~\ref{table:cg_crashes}). Because of this, \cg
even achieved similar or lower throughput than \sa for W6 and W7 in P100s 
and W3 and W6 in V100s. W1 in V100s is an exceptional case where the \cg 
scheduler managed to run efficiently without crashing, 
leading to higher throughput than \shortname. W1 has a
1:1 ratio of large:small jobs, and on this workload, \cg was (coincidentally) 
able to do a better packing of the job mix without crashing.
In general, Table~\ref{table:cg_crashes} shows that the crash behavior of
the \cg scheduler was erratic (and increasingly so as the number of workers 
increased). In particular, for job mixes with large jobs, the percentage 
of crashes due to \cg is alarming, ranging from 13\% to 50\% on V100s. 
This implies it is unlikely to be useful in practice 
unless, for example, workload sizes are known and guaranteed not 
to overflow the memory capacity of the GPUs.

\begin{table}[!tbp]
 \centering
  \caption{Percentage of crashed jobs for \cg (P100s/V100s)}
 \begin{tabular}{|c|l|l|l|l|}
  \hline
  \textbf{\# of workers} & \bf 1:1 mix & \bf 2:1 & \bf 3:1 & \bf 5:1  \\
  \hline
  \hline
    3/6	 &	$0$/$0$       & $3\%$/$6\%$   &	$8\%$/$17\%$  &	$0$/$0$ \\
    4/8	 &	$14\%$/$13\%$ & $6\%$/$19\%$  &	$6\%$/$25\%$  &	$9\%$/$13\%$ \\
    5/10	 &	$13\%$/$15\%$ & $13\%$/$25\%$ &	$20\%$/$20\%$ &	$22\%$/$25\%$ \\
    6/12	 &	$16\%$/$33\%$ & $17\%$/$29\%$ &	$16\%$/$38\%$ &	$16\%$/$50\%$  \\
  \hline
  \end{tabular}
  \label{table:cg_crashes}
\end{table}



\subsection{Turnaround Time Speedup}
As mentioned, the experiment begins with a queue already full of jobs.
We view these jobs as requests, and measure the turnaround time for each 
job. While some degree of slowdown can happen when a particular job is 
co-executing with others, the turnaround time (time interval 
between the job arrival time and completion time) can be boosted 
by improving the throughput and reducing the time these requests sit in 
the queue. Table \ref{table:mgb_avg_turnaround} shows the turnaround time
speedups over \sa for all mixes and workload sizes on both the P100s
and V100s. We observed an average of 3.7x for the P100s and 2.8x for
the V100s, and a maximum of almost 5x in some cases.

\begin{table}
 \centering
  \caption{\shortname average turnaround speedup}
 \begin{tabular}{|l|l|l|l|l|l|}
  \hline
  \bf GPUs & \textbf{\# of jobs} & \bf 1:1 mix & \bf 2:1 & \bf 3:1 & \bf 5:1  \\
  \hline
  \hline
    2xP100s & 16 jobs	& 4.9x	& 2.3x	& 4.9x	& 4.3x \\
    2xP100s & 32 jobs	& 4.6x	& 3.2x	& 3.6x	& 2.0x \\
    4xV100s & 16 jobs	& 2.4x & 2.0x	& 3.5x	& 2.6x \\
    4xV100s & 32 jobs	& 3.8x & 2.9x	& 2.9x	& 2.6x \\
  \hline
  \end{tabular}
  \label{table:mgb_avg_turnaround}
\end{table}

\begin{figure}[!tbp]
    \centering
    \includegraphics[width=\columnwidth]{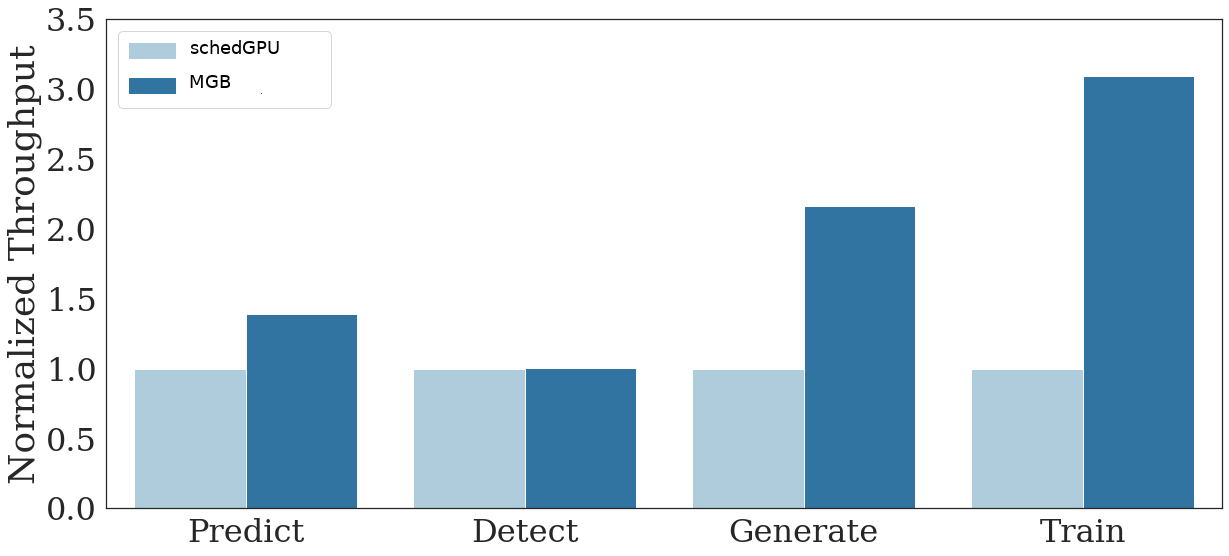}
    \caption{Throughput on homogeneous, parallel 8-job neural network workloads on 4xV100s}
    \vspace{-.1in}
    \label{figure:8-job-darknet-chao}
\end{figure}

\subsection{Neural Network Performance} \label{sec:neural}
We ran two neural network experiments. In the first we compared
\shortname against schedGPU ~\cite{RSNV2018},
a state-of-the-art work for intra-node scheduling.
We ran 4 homogeneous workloads
(1 for each type of task described in section \ref{sec:workloads}), with 8 jobs in each workload.
On our 32-core, 4xV100 AWS system, this corresponds to 1 out of every
4 CPU cores creating work for the GPU.
Thus, the system is not underloaded nor overloaded.
Further, each task's network is between 0.5-1.5GB, so 8 jobs can always
fit within a single V100's memory. For this reason, we faithfully
mimic schedGPU, which would schedule all jobs to run on one device,
since the memory capacity is not exceeded. (Note that
schedGPU uses memory capacity as the only resource criterion in scheduling;
please see the related work section for more details).

Figure~\ref{figure:8-job-darknet-chao} shows the results for schedGPU
and \shortname.
schedGPU does not handle device reassignment, and 
it only ensures memory capacity is not exceeded.
Thus, it underperforms on modern neural network loads
such as used in these experiments, which have a high compute resource need in terms of warps or
thread blocks.
Because it does not account for this resource need, schedGPU is unable to
spread work across GPUs, and could oversaturate a GPU. 
\shortname achieves throughput speedups of 1.4x, 2.2x, and 3.1x
over schedGPU for the predict,
generate, and train tasks, respectively. For detection, the frameworks
have similar results. Real-time object detection networks are
designed for video streaming
and must be performant at runtime. The network we tested can
process 244 FPS on the older, NVIDIA Pascal Titan X. In fact,
the nvidia-smi tool reports 25\% or less for volatile GPU utilization
on these tasks, so the compute units are not saturated in this case.
One takeaway is that single-GPU performance, even when it satisfies the
simultaneous memory requirements of all running jobs, can and will suffer
under common, modern machine learning tasks.
A second is that memory requirements alone, even
when multiple GPUs are used to separate jobs, misses compute requirements (among
others) that determine performance.

Finally, we ran one large-scale experiment in a manner similar to our
Rodinia setup, in order to verify that \shortname is effective
on large mixes of these neural network jobs.
We ran a 128-job, random mix of the 4 tasks.
With 32 workers, \shortname completed the jobs 2.7x faster than
single-assignment, which is comparable to the results we see for Rodinia.

\subsection{Kernel Slowdown}
We looked closely at the kernel slowdowns, i.e. the amount of extra
time to run a given kernel on the GPU.
We include Algorithm~\ref{alg:schedule} in this study. Despite worse
throughput than Algorithm~\ref{alg:schedule2}, we expected its kernel slowdown
to be an improvement (because it may hold back jobs based on
its exact knowledge of compute resources).
The results were interesting.
We compared the two algorithms to the baseline single-assignment
performance. On the same 8 workloads (Table \ref{tab:workloads})
and 4xV100 system, Algorithm~\ref{alg:schedule} averaged 1.8\%,
whereas Algorithm~\ref{alg:schedule2} averaged 2.5\% (Table~\ref{tab:kernel_slowdowns}). 
Thus, both algorithms cause
negligible slowdowns to the kernels themselves; and compared with
each other, the difference is less than 1\%.

\begin{table}
 \centering
  \caption{Kernel slowdowns for Algorithm~\ref{alg:schedule} and Algorithm~\ref{alg:schedule2} on 8 workloads
           on 4xV100s, expressed as a percentage of single-assignment performance.}
  \hspace{-0.1pt}
 \begin{tabular}{|l|l|l|l|l|l|l|l|l|l|}
  \hline
  \bf Sched & \bf W1 & \bf 2 & \bf 3 & \bf 4 & \bf 5 & \bf 6 & \bf 7 & \bf 8 & \bf Avg  \\
  \hline
  \hline
         Alg2 & -0.3 & 1.0 & 0.3 & 4.1 & 2.9 & 5.1 & 1.1 & 0.6 & 1.8 \\
         Alg3 & -0.7 & 0.8 & 7.0 & 3.1 & 2.2 & 4.1 & 0.4 & 2.9 & 2.5 \\
  \hline
  \end{tabular}
  \label{tab:kernel_slowdowns}
\end{table}

\section{Related Work} \label{sec:related}
The importance of dealing with GPU sharing among different workloads 
is widely recognized in recent research. To the best of our knowledge,
this work is the first 
to tackle this problem for multi-GPU systems in a fully automated manner 
with no user intervention,
but its design and 
prototype are inspired by many state-of-the-art studies. In this 
section, we present a brief survey of the most related studies.


As mentioned, when a node has multiple GPUs, processes
are still, by default, automatically assigned to $device0$. 
Prior works either have no visibility
into each process' device selection, or they do not have
any control over it.
For example, Slurm \cite{YoJG2003} can effectively manage
job queues and ensure that when an independent job runs on a node, the node
is provisioned with a sufficient number of (available) GPUs for that job.
Slurm has no way, however, of scheduling jobs to specific GPUs within a node.
Consider two independent jobs running on a 2-GPU node, both of which
need only one GPU but are defaulting to $device0$. Slurm
has no way of assigning one job to $device0$ and the other to $device1$.
Slurm would unnecessarily sequential jobs.

In the work by Reaño et al. \cite{RSNV2018},
schedGPU schedules multiple kernels on the same
GPU without overrunning the memory.
It differs from \shortname in several ways.
First, schedGPU requires the programmer to add library calls
that pass the applications' memory needs, and this can be error prone.
Estimating the memory needs
of complex applications can be a daunting task, and in fact we had to
resort to solutions such as a lazy runtime
to accurately bind the memory calls to a CUDA task.
Second, it takes into account only memory, which, as we show in a simple
neural network experiment, can suffer from slowdowns if compute is
not properly managed both within and across GPUs (and which our algorithms
take into account).
Lastly, schedGPU is designed for a single-device environment,
and only has the capability of suspending or continuing a CUDA
operation. Solving this automatically is non-trivial and requires the
ability to make a CUDA task device-independent and to handle
device reassignment.
It requires a system that can identify CUDA tasks and map CUDA
operations within tasks to appropriate 
devices based on the devices' statuses.
This is a critical feature required 
for scheduling on multi-GPU systems and is provided by our framework.
Due to this reason, schedGPU
is inapplicable to modern systems with 2, 4, or more GPUs
per node, which are essential
to meet the requirements of modern machine learning loads.

Numerous frameworks are designed specifically for deep learning.
Gandiva \cite{XBRS2018}
solves a cluster-wide GPU scheduling
problem for frameworks like PyTorch~\cite{pytorch}
and TensorFlow~\cite{tensorflow}
by exploiting properties of deep learning (e.g. prioritizing
certain jobs based on feedback-driven exploration).
Amaral et al. \cite{DBLP:conf/sc/AmaralPCSS17}
have a topological perspective (considering GPU count and placement,
but also the network connections between them); the design is quite different,
incorporating, for example, ``service-level objectives'' that an application
must express before it can be mapped effectively onto a topology.
\citeauthor{HRKG2018}~\cite{HRKG2018} consider the problem of scheduling
multiple concurrent deep neural networks in a server system on a 
single GPU.
MXNet~\cite{mxnet} partitions data batches among GPUs,
and each GPU is dedicated to the assigned workload (a process).
It does not have a scheduler and cannot train multiple models simultaneously
on a given device to improve resource utilization when a single model 
training cannot saturate the device.
In contrast, \shortname is designed to handle this case. With \shortname,
multiple models
can be trained simultaneously on the same set of devices, thereby improving
the resource utilization and the system throughput. More importantly,
\shortname targets any general GPU workload (ML workloads being
a special case), which distinguishes it from these prior works
We also focus on the problem of GPU 
sharing inside a node, rather than in a cluster. In addition our scheme
is fully automated with compiler analysis in the driver's seat,
leading to memory safety and other performance guarantees. 

Other prior art is focused on different problems.
FLEP \cite{WLZJ2017} tackles the problem of how to do effective
preemption to regain resources such as SMs for scheduling
higher priority processes. In contrast, we solve a different
problem related to how to pack GPUs effectively given the knowledge of their resource needs. Solutions such as FLEP
can be coupled with ours to yield resources which can then be effectively
shared using our system. FLEP does not address the issue of 
device placement.
CODA \cite{DBLP:journals/taco/KimHNKJEKL18} solves
the problem of where to place data in a multi-GPU system.
Several frameworks~\cite{BaKa2012,SaWB2013,ZhTL2015, WLZJ2017} have 
been proposed to enable preemption on GPUs through kernel slicing. 
They slice long-running kernel invocations into multiple short-running 
sub-kernels. This allows GPU applications to be preempted when sub-kernel 
invocations are ﬁnished. \citeauthor{TGCR2014}~\cite{TGCR2014,PaPM2015} 
proposed hardware extensions to enable the preemption. However, these 
frameworks are orthogonal to our work, are mainly designed for
single-GPU systems and do not solve 
the multi-tenancy, multi-GPU problem we tackle. They can however be 
integrated with our scheduling framework to improve QoS support for 
latency-critical applications. 
\citeauthor{CYMT2016}~\cite{CYMT2016} orchestrate
user-facing applications and throughput-oriented applications to improve 
the GPU utilization while maintaining the QoS.


Gdev~\cite{KMMB2012} integrates runtime support for GPUs into the OS and 
provides ﬁrst-class GPU resource management schemes for multitasking systems. 
PTask~\cite{RCSR2011} is another approach that makes the OS GPU-aware. It uses 
a task-based data flow programming model and exposes the task graph to the 
OS. These methods would require significant changes to basic system 
software stacks; in contrast, our solution offers full automation 
with no changes to an application or any part of the GPU software stack
by providing a user-level scheduler.

NVIDIA's unified memory allows for
memory oversubscription, but it also incurs significant performance
overheads when data has to be migrated to or from the device.
It can be convenient, but when transfer overheads cannot be hidden, it
goes against the practice of high performance
programming. None of our benchmarks have any instance of its usage.
NVIDIA's multi-instance GPU (MIG) is a new partitioning and isolation feature.
MIG does not solve the problem of packing independent processes across multiple
GPUs, but it could be interesting future work to consider it in our framework.
\section{Conclusion}\label{sec:concl}
In this paper, we present a fully automated GPU scheduling framework 
to uniformly and transparently manage GPU resources. It constructs 
CUDA tasks via static program analysis coupled with a lazy
runtime, and schedules CUDA tasks from independent workloads onto 
GPU devices. It leverages the compiler to insert 
probes for CUDA tasks, and these probes gather the resource 
requirements of tasks at runtime and convey them to a user-level 
scheduler. The scheduler places tasks onto devices based on their 
resource requirements and the devices' statuses. With the knowledge 
of resource requirements for each CUDA task, it guarantees memory 
safety among co-executing tasks from independent processes (i.e. 
no crashes due to insufficient space). We evaluated the system on 
the Rodinia benchmark suite on two different GPU
families - Pascal and Volta. On average, on a 
2-GPU Pascal (P100) system, the new scheduling framework improves 
system throughput by 2.2$\times$ over a memory-safe scheduler, and 
by 64$\%$ over a memory-unsafe scheduler that has a crash frequency 
of 11$\%$. On a 4-GPU Volta (V100) system, the throughput improves by 
an average of 2$\times$ over a memory-safe scheduler, and by 41$\%$ 
over a memory-unsafe scheduler with a crash frequency of $20\%$.
We also evaluated the 4-GPU system on neural network workloads and
measured similar results (2.7$\times$ throughput improvement over the memory-safe
scheduler). Over a competing state of the art GPU scheduling technique ~\cite{RSNV2018}, \shortname shows
a throughput improvement ranging from 1.4$\times$ to 3.1$\times$.  Two scheduling
algorithms are presented which leverage MPS and yield very small degradation
of 1.8\% and 2.5\% to individual kernels' execution speed. Supported by this
empirical evaluation, we 
believe that such an automated solution is a practical way of 
solving the GPU sharing problem to boost throughput and device utilization. 

\bibliographystyle{IEEEtranN}
\bibliography{refs.bib}


\end{document}